\documentclass[12pt]{article}

\normalsize
\input{epsf}

\begin{document}
\addtolength{\baselineskip}{.35mm}
\newlength{\extraspace}
\setlength{\extraspace}{2.5mm}
\newlength{\extraspaces}
\setlength{\extraspaces}{2.5mm}

\newcommand{\newsection}[1]{
\vspace{15mm} \pagebreak[3] \addtocounter{section}{1}
\setcounter{subsection}{0} \setcounter{footnote}{0}
\noindent {\Large\bf \thesection. #1} \nopagebreak
\medskip
\nopagebreak}

\newcommand{\newsubsection}[1]{
\vspace{1cm} \pagebreak[3] \addtocounter{subsection}{1}
\addcontentsline{toc}{subsection}{\protect
\numberline{\arabic{section}.\arabic{subsection}}{#1}}
\noindent{\large\bf %\thesection.
\thesubsection. #1} \nopagebreak \vspace{3mm} \nopagebreak}
\newcommand{\ba}{\begin{eqnarray}
\addtolength{\abovedisplayskip}{\extraspaces}
\addtolength{\belowdisplayskip}{\extraspaces}
\addtolength{\abovedisplayshortskip}{\extraspace}
\addtolength{\belowdisplayshortskip}{\extraspace}}

\newcommand{\be}{\begin{equation}
\addtolength{\abovedisplayskip}{\extraspaces}
\addtolength{\belowdisplayskip}{\extraspaces}
\addtolength{\abovedisplayshortskip}{\extraspace}
\addtolength{\belowdisplayshortskip}{\extraspace}}
\newcommand{\ee}{\end{equation}}
\newcommand{\STr}{{\rm STr}}
\newcommand{\figuur}[3]{
\begin{figure}[t]\begin{center}
\leavevmode\hbox{\epsfxsize=#2 \epsffile{#1.eps}}\\[3mm]
\parbox{15.5cm}{\small
\it #3}
\end{center}
\end{figure}}
\newcommand{\im}{{\rm Im}}
\newcommand{\calm}{{\cal M}}
\newcommand{\call}{{\cal L}}
\newcommand{\sect}[1]{\section{#1}}
\newcommand\hi{{\rm i}}
\def\bea{\begin{eqnarray}}
\def\eea{\end{eqnarray}}

\begin{titlepage}
\begin{center}

\vspace{3.5cm}

{\Large \bf{Shear Viscosity from AdS Born-Infeld Black Holes}}\\[1.5cm]

{Rong-Gen Cai $^{a,}$\footnote{Email: cairg@itp.ac.cn},} {Ya-Wen Sun
$^{a,b,}$\footnote{Email: sunyw@itp.ac.cn},}

\vspace*{0.5cm}

{\it $^{a}$ Institute of Theoretical Physics, \\Chinese Academy of
Sciences P.O.Box 2735, Beijing 100190, China

$^{b}$Graduate University of Chinese Academy of Sciences,
\\ YuQuan
Road 19A, Beijing 100049, China}

\date{\today}% It is always \today, today,

\vspace{3.5cm}

\textbf{Abstract} \vspace{5mm}

\end{center}
We calculate the shear viscosity in the frame of AdS/CFT
correspondence for the field theory with a gravity dual of
Einstein-Born-Infeld gravity. We find that the ratio of $\eta/s$ is
still the conjectured universal value $1/4\pi$ at least up to the
first order of the Born-Infeld parameter $1/b^2$.

\end{titlepage}

\newpage

\section{Introduction}
The AdS/CFT
correspondence~\cite{Maldacena:1997re,Gubser:1998bc,Witten:1998qj,Aharony:1999ti}
has been a useful way to calculate dynamical quantities of strongly
coupled gauge theories. A famous example is the discovery of the
universality of the ratio of the shear viscosity $\eta$ to the
entropy density $s$, which is equal to $1/4\pi$ in all theories in
the regimes described by gravity
duals~\cite{Policastro:2001yc,Kovtun:2003wp,Buchel:2003tz,Kovtun:2004de,{Buchel:2004qq}}.
This ratio is also conjectured to be a universal lower bound (the
KSS bound) for all materials~\cite{Kovtun:2003wp}. All known
materials in nature by now satisfy this bound.
In~\cite{Mas:2006dy,{Son:2006em},{Saremi:2006ep},{Maeda:2006by}},
the authors also calculated the shear viscosity of gauge theories
with chemical potentials turned on by studying R-charged black
holes. With the presence of nonzero chemical potentials the ratio of
shear viscosity to entropy density is still $1/4\pi$. Also the
stringy corrections to the ratio were calculated in
\cite{{Buchel:2004di},Benincasa:2005qc,{Buchel:2008ac},{Buchel:2008wy},{Buchel:2008sh},{Myers:2008yi}}
where the corrections to the value $1/4\pi$ are found to be positive
and satisfy the lower bound. More discussions on this KSS bound can
be found in
\cite{{Policastro:2002se},{Policastro:2002tn},{Cohen:2007qr},{Cherman:2007fj},
{Chen:2007jq},{Son:2007xw},{Fouxon:2008pz},{Dobado:2008ri},{Landsteiner:2007bd}}.

On the other hand, in~\cite{Brigante:2007nu,{KP}} the authors
considered  $R^2$ corrections in the gravity side and found that the
modification of the ratio of shear viscosity over entropy density to
the conjectured bound is negative, which means that the lower bound
could be violated in that case. The higher derivative gravity
corrections they considered can be seen as generated from stringy
corrections given the vastness of the string landscape. They gave a
new lower bound, ${4}/{25\pi}$, based on the causal condition.
However, the physical implication of this violation of the bound is
still not very clear yet.

This motivates us to consider whether higher derivative corrections
to the gauge matter fields on the gravity side also affect the value
of shear viscosity when chemical potentials are turned on. In
\cite{Mas:2006dy,{Son:2006em},{Saremi:2006ep},{Maeda:2006by}}, the
ratio of $\eta/s$ for the case of nonzero chemical potential was
calculated through Einstein-Maxwell theory on the gravity side. As
an example of nonlinear electrodynamics, we consider
Einstein-Born-Infeld theory with a negative cosmological constant.
This theory can be viewed as a nonlinear extension on the gauge
fields in Einstein-Maxwell theory.

In this paper we will calculate the shear viscosity of gauge
theories with the gravity dual of Einstein-Born-Infeld theory. The
background we need is just the AdS Born-Infeld black hole solution
\cite{Dey:2004yt,Cai:2004eh}. Calculate the shear viscosity in this
background and we can see if the higher derivative corrections to
the matter fields which are coupled to gravity will also affect the
ratio. And the answer we get is that the ratio is not affected by
this higher derivative correction to the gauge fields at least at
the first non-trivial order. This fact along with the modification
of the ratio in higher derivative gravity theories may imply that
the ratio may be mainly determined by the form of the action of
gravity in the dual gravity description. Here the action should be
viewed as the effective action which includes the contribution from
stringy corrections. If we do not consider the stringy corrections,
the effective action for gravity is just the Einstein-Hilbert
gravity term. In the case of Einstein-Hilbert gravity, this has been
verified in various cases. However, in Gauss-Bonnet gravity the
ratio of $\eta/s$ has a correction when gauge fields are coupled to
gravity~\cite{Ge:2008ni}.

In this paper, we first present some basic properties of AdS
Born-Infeld black holes in Sec.~2.  We calculate the shear viscosity
of gauge theories with the gravity dual of Einstein-Born-Infeld
theory to the first order of the Born-Infeld parameter $1/b^2$
through Kubo-formula in Sec.~3. Sec.~4 is devoted to conclusions and
discussions.

\section{AdS Born-Infeld black holes}

In this section we give some basic properties of the Born-Infeld
black hole solution in the presence of a negative cosmological
constant in five dimensions. The action can be written as
%\begin{equation}\end{equation}
\begin{equation}
S=\frac{1}{16\pi G}\int d^5x\sqrt{-g}\Big( R-2\Lambda+L(F) \Big),
\end{equation}
where $L(F)=4b^2(1-\sqrt{1+\frac{F_{\mu\nu}F^{\mu\nu}}{2b^2}})$. The
constant $b$ here is the Born-Infeld parameter and has the dimension
of mass. In the limit of $b\rightarrow \infty$, $L(F)$ reduces to
the Maxwell form. Thus if we expand $L(F)$ in a series of $1/b^2$,
we will find that the $1/b^2$ corrections to the Maxwell action just
correspond to the higher derivative corrections of the gauge fields.

We can write out the equations of motion explicitly as
\begin{equation}\label{equ}
R_{\mu\nu}-\frac{1}{2}g_{\mu\nu}R+\Lambda
g_{\mu\nu}-\frac{1}{2}g_{\mu\nu}L(F)-\frac{2F_{\mu\lambda}F_{\nu}^{~\lambda}}
{\sqrt{1+\frac{F^2}{2b^2}}}=0,
\end{equation}
and
\begin{equation}
{\bigtriangledown}_{\mu}\bigg(\frac{F^{\mu\nu}}{\sqrt{1+\frac{F^2}{2b^2}}}\bigg)=0.
\end{equation}
Since we want to calculate the shear viscosity of the corresponding
field theory living in $R^{1,3}$, we need a black hole solution with
a Ricci-flat horizon. The AdS Born-Infeld black hole solution  with
a Ricci-flat horizon can be written as~\cite{Cai:2004eh}
\begin{equation}
\label{eq4}
 ds^2=-V(r)dt^2+\frac{1}{V(r)}dr^2+r^2(d\vec{x}^2),
\end{equation}
where
\begin{equation}
V(r)=-\frac{m}{r^2}+(\frac{b^2}{3}+\frac{1}{l^2})r^2-\frac{b}{3r}\sqrt{b^2r^6+3q^2}
+{\frac{3q^2}{2r^4}}{}_2F_1[\frac{1}{3},\frac{1}{2},\frac{4}{3},-\frac{3q^2}{b^2r^6}],
\end{equation}
and the only nonzero component of $F_{\mu\nu}$ is
\begin{equation}
F^{rt}=\frac{\sqrt{3}bq}{\sqrt{b^2r^6+3q^2}}.
\end{equation}
Here $q$ is an integration constant which is related to the electric
charge of the black hole and $l$ is the AdS radius through
$\Lambda=-6/l^2$. When $b^2$ approaches to infinity, this solution
becomes the AdS Reissner-Nordstr\"om black hole solution.

We can get the position of the outer horizon by solving $V(r_+)=0$.
For future convenience we define $u={r_+^2}/{r^2}$ and rescale
$\vec{x}$ to the new coordinate system
\begin{equation}
ds^2=-V(u)dt^2+\frac{r_+^2}{4u^3V(u)}du^2+\frac{r_+^2}{ul^2}d\vec{x}^2.
\end{equation}
Now $V(u)$ becomes
\begin{equation}
V(u)=-\frac{mu}{r_+^2}+(\frac{b^2}{3}+\frac{1}{l^2})\frac{r_+^2}{u}
-\frac{b}{3r_+}\sqrt{\frac{b^2r_+^6}{u^2}+3q^2u}+\frac{3q^2u^2}{2r_+^4}
{}_2F_1[\frac{1}{3},\frac{1}{2},\frac{4}{3},-\frac{3q^2u^3}{b^2r_+^6}],
\end{equation}
and the horizon corresponds to $u=1$. Then the mass parameter $m$
can be expressed by $r_+$ as
\begin{equation}
m=\left(\frac{b^2}{3}+\frac{1}{l^2} \right
)r_+^4-\frac{br_+}{3}\sqrt{{b^2r_+^6}+3q^2}+\frac{3q^2}{2r_+^2}
{}_2F_1[\frac{1}{3},\frac{1}{2},\frac{4}{3},-\frac{3q^2}{b^2r_+^6}].
\end{equation}
The thermodynamic properties of this black hole has been discussed
in~\cite{Dey:2004yt,{Cai:2004eh},{Fernando:2006gh},{Sheykhi:2006dz},{Miskovic:2008ck},{Myung:2008eb}}.
Here we only give the entropy density of this black hole solution
for future use
\begin{equation}
s=\frac{1}{4G} \frac{r_+^3}{l^3}.
\end{equation}

\section{Shear viscosity from AdS Born-Infeld black holes}

In this section we calculate the shear viscosity of the field
theory dual to the black hole  background (\ref{eq4})  through the
Kubo-formula~\cite{Policastro:2002se,{Son:2007vk}}:
%\begin{equation} \end{equation}
\begin{equation}\label{eta}
\eta=\lim_{\omega\rightarrow 0}\frac{1}{2\omega
i}\Big(G^A_{xy,xy}(\omega,0)-G^R_{xy,xy}(\omega,0)\Big),
\end{equation}
where $\eta$ is the shear viscosity, and the retarded Green's
function is defined by
\begin{equation}
G^R_{\mu\nu,\lambda\rho}(k)=-i\int d^4xe^{-ik\cdot x}\theta (t)
\langle[T_{\mu\nu}(x),T_{\lambda\rho}(0)] \rangle.
\end{equation}
The advanced Green's function can be related to the retarded Green's
function
by$G^A_{\mu\nu,\lambda\rho}(k)=G^R_{\mu\nu,\lambda\rho}(k)^{*}$. We
compute the retarded Green's function by making a small perturbation
of graviton. Here we choose spatial coordinates so that the momentum
of the perturbation points along the $z$-axis. Then the
perturbations can be written as $h_{\mu\nu}=h_{\mu\nu}(t,z,r)$. In
this basis there are three groups of gravity perturbations, each of
which is decoupled from others: the scalar, vector and tensor
perturbations~\cite{Kovtun:2005ev}. Here we use the simplest one,
the tensor perturbation $h_{xy}$. We use $\phi$ to denote this
perturbation $\phi=h^x_y$ and write $\phi$ in a basis as
$\phi(t,u,z)=\phi(u)e^{-i\omega t+ip z}$.

We can get the equation of motion of this $\phi(u)$ by perturbing
both sides of the equation of motion (\ref{equ}) to the first
order of $\phi(u)$
\begin{equation}\label{equation}
\phi''(u)+A\phi'(u)+B\phi(u)=0.
\end{equation}
Here to avoid complication we only calculate the shear viscosity
up to the first order in the parameter $1/b^2$, and in this
approximation we have
\begin{equation}
A=A_0+A_1,
\end{equation}
where $A_0$ denotes the part of the coefficient $A$ in the limit
$b^2\rightarrow\infty$,
\begin{equation}
A_0=\frac{l^2q^2(1-2u)u^2+r_+^6(1+u^2)}{u(u-1)(r_+^6(1+u)-l^2q^2u^2)},
\end{equation}
while $A_1$ denotes the part of the first order correction of
$1/b^2$:
\begin{equation}
A_1=\frac{3l^2q^4u(2r_+^6(1+u)^2(1+2u^2)-l^2q^2u^3(1+2u+3u^2))}{16b^2r_+^6(l^2q^2u^2-r_+^6(1+u))^2}.
\end{equation}
Also $B$ can be written as the sum of two parts,
%\begin{equation}\end{equation}
\begin{eqnarray}
&&B =B_0+B_1 \nonumber
\\
&&~~~=\frac{r_+^6({\bar{p}}^2(u-1)(r_+^6(1+u)-l^2q^2u^2)+r_+^6{\bar{w}}^2)}{u(u-1)^2(r_+^6(1+u)-l^2q^2u^2)^2}
\nonumber
\\&&~~~+\frac{3l^2q^4u(1+u+u^2+u^3)(\bar{p}^2(u-1)(-r_+^6(1+u)+l^2q^2u^2)-2r_+^6\bar{w}^2)}{16b^2(u-1)^2(r_+^6(1+u)-l^2q^2u^2)^3}.
\end{eqnarray}
Here $\bar{w}={l^2\omega}/{2r_+}$ and $\bar{p}={l^2p}/{2r_+}$.
$A_1$ and $B_1$ manifest the contribution of the higher derivative
corrections to the gauge field on the gravity side. To solve for
$\phi(u)$, we write
\begin{equation}\label{phi}
\phi(u)=(1-u)^{-i\beta\bar{w}}F(u)
\end{equation}
to decide the boundary condition near the horizon, where $\beta$
is a constant to be fixed. Substituting (\ref{phi}) into
(\ref{equation}) and solving it near the horizon $u=1$, we get
\begin{equation}
\beta=\beta_0+\beta_1=\frac{r_+^6}{2r_+^6-l^2q^2}-\frac{3l^2q^4}{4b^2(2r_+^6-l^2q^2)^2},
\end{equation}
by pure incoming wave boundary condition near $u=1$.  Here
$\beta_1$ is the contribution of the higher derivative correction.
Next we move on to solve $\phi(u)$ in the whole spacetime. Because
 we know from (\ref{eta}) that we only need the low frequency
behavior of $\phi(u)$ to calculate the shear viscosity, we can
expand $F(u)$ in a power series of $\bar{w}$ and $\bar{p}$
\begin{equation}
F(u)=1+i\beta_0\bar{w}F_0(u)+i\beta_1\bar{w}F_1(u)+O(\bar{w}^2)+O(\bar{p}^2).
\end{equation}
In this expansion, $F_1$ represents the contribution from the
higher derivative correction of the gauge field. The equations of
motion of $F_0(u)$ and $F_1(u)$ can be derived at the first order
of $\bar{w}$ separately,
\begin{equation}
F_0''(u)+A_0F_0'(u)+\frac{1}{(1-u)^2}+\frac{A_0}{1-u}=0,
\end{equation}
and
\begin{equation}
F_1''(u)+A_0F_1'(u)+\frac{\beta_0}{\beta_1}A_1F_0'(u)+\frac{\beta_0A_1}{\beta_1(1-u)}+\frac{1}{(1-u)^2}+\frac{A_0}{1-u}=0.
\end{equation}
The solutions of these two linear differential equations can be
uniquely decided with the boundary condition
$F_0(u)|_{u=0}=F_1(u)|_{u=0}=0$ and the constraint that both
$F_0(u)$ and $F_1(u)$ should be regular at the horizon $u=1$. The
solutions are
\begin{equation}
F_0(u)=\frac{1}{2}(\ln\frac{r_+^6(1+u)-l^2q^2u^2}{r_+^6}-\frac{3r_+^3}{{\sqrt{4l^2q^2+r_+^6}}}
\ln{\frac{u\sqrt{4l^2q^2+r_+^6}+r_+^3(2+u)}{-u\sqrt{4l^2q^2+r_+^6}+r_+^3(2+u)}}),
\end{equation}
and
\begin{eqnarray}
\nonumber
F_1(u)&=&C_0(u)+C_1\ln{\frac{(r_+^3\sqrt{4l^2q^2+r_+^6}+r_+^6-2l^2q^2u)
(\sqrt{4l^2q^2+r_+^6}-r_+^3)}{(r_+^3\sqrt{4l^2q^2+r_+^6}-r_+^6+2l^2q^2u)(\sqrt{4l^2q^2+r_+^6}+r_+^3)}}\\
&+&C_2\ln{\frac{r_+^6(1+u)-l^2q^2u^2}{r_+^6}},
\end{eqnarray}
where $C_0(u)$ is a function of $u$,
\begin{eqnarray}
\nonumber C_0(u)=&-&\frac{(l^2q^2 - 2r_+^6)u}{8l^6q^6(4l^2q^2 +
r_+^6)(l^2q^2u^2 - r_+^6(1 + u))}\times\\ \nonumber
&&\bigg(12r_+^{24}(1 + u) - 6l^2q^2r_+^{18}(-9 -
10u + u^2) + 4l^8q^8u(1 + 4u + u^2) +\\
     &&l^4q^4r_+^{12}(30 + 51u - 29u^2 - 2u^3) - l^6q^6r_+^6(6 + 2u + 16u^2 +
     7u^3)\bigg),
\end{eqnarray}
$C_1$ and $C_2$ are two constants,
\begin{eqnarray}
&& C_1=\frac{3(10l^{10}q^{10}r_+^3+ 25l^8q^8r_+^9 +
12l^6q^6r_+^{15} - 45l^4q^4r_+^{21} - 28l^2q^2r_+^{27} -
4r_+^{33})}{8l^8q^8(4l^2q^2+r_+^6)^{3/2}}  \\
&& C_2=\frac{l^8q^8 + 6l^6q^6r_+^6 + 9l^4q^4r_+^{12} -
12l^2q^2r_+^{18} - 12r_+^{24}}{8l^8q^8}.
\end{eqnarray}
Now we want to get the on-shell action for the perturbation. The
on-shell action for $\phi(u)$ is a sum of two parts: one is from the
bulk action $S_{bulk}$ and the other from the Gibbons-Hawking
boundary term $S_{GH}$. We first expand
\begin{equation}
\phi(x,u)=\int \frac{d^4k}{(2\pi)^4}e^{ikx}f(k)\phi_k(u).
\end{equation}
The bulk action for this perturbation then is
\begin{equation}
S_{bulk}=\frac{1}{16\pi G}\int
\frac{d^4k}{(2\pi)^4}f(k)f(-k)\int_{1}^{0}du(K_{1}\phi''_k\phi_{-k}+K_{2}\phi'_k\phi'_{-k}+K_{3}\phi_k'\phi_{-k}+K{_4}\phi_k\phi_{-k})
\end{equation}
where $K_{1}=2\sqrt{-g}g^{uu}$ and
$K_{2}=\frac{3}{2}\sqrt{-g}g^{uu}$.  The $K_3$ and $K_4$ terms are
not relevant to our aim, so we do not explicitly write them here.
The Gibbons-Hawking boundary term is
\begin{equation}\label{GB}
S_{GH}=\frac{1}{8\pi}\int_{\partial M} d^4x\sqrt{-h}K.
\end{equation}
Substituting the solution of $\phi(u)$ into (\ref{GB}) and we
reach
\begin{equation}
S_{GH}=\frac{1}{16\pi}\int
\frac{d^4k}{(2\pi)^4}f(k)f(-k)(K_5\phi_k\phi_{-k}+K_{6}\phi'_k\phi_{-k}),
\end{equation}
where $K_6=-2\sqrt{-g}g^{uu}$ and $K_5$ only contributes to the
Green's function a real part thus not relevant to the following
calculations.

By applying the equation of motion (\ref{equation}), the bulk
action becomes  two surface terms
\begin{equation}
S_{bulk}=\frac{1}{16\pi G}\int \frac{d^4k}{(2\pi)^4}f(k)f(-k)
\bigg(\int_1^0du[E.O.M]\phi_{-k}+(\frac{K_3-K_1'}{2}\phi_k\phi_{-k}+K_2\phi_k'\phi_{-k})|_1^0\bigg).
\end{equation}
The Gibbons-Hawking term is itself a boundary contribution, so the
total action can be written as
\begin{equation}\label{33}
S=\int \frac{d^4k}{(2\pi)^4}f(k)f(-k)F(k,u)|_0^1,
\end{equation}
where $F(k,u)$ can be expressed by the constants $K_i$. The
retarded Green's function can be calculated in the way
\begin{equation}\label{34}
G^R_{xy,xy}(k)=-2F(k,u=0).
\end{equation} We substitute the solution into (\ref{33})
and can obtain
\begin{equation}
F(k,u=0)=\frac{1}{32\pi G}\sqrt{-g}g^{uu}\phi'_k\phi^*_{k}|_{u=0}
=\frac{1}{16\pi G}i\bar{w}\frac{r_+^4}{l^5}=\frac{1}{16\pi
G}i\omega\frac{r_+^3}{2l^3}.
\end{equation}
Substituting this into (\ref{34}) and (\ref{eta}), we have
\begin{equation}\label{result}
\eta=\lim_{\omega\rightarrow 0}\frac{F-F^*}{wi}=\frac{1}{16\pi
G}\frac{r_+^3}{l^3}.
\end{equation}
Comparing this with $s=\frac{1}{4G}\frac{r_+^3}{l^3}$,  we finally
reach $\eta/s=1/4\pi$.

Here we only performed the calculation to the first order of the
parameter $1/b^2$. That is to say, we considered the effect on the
ratio of $\eta/s$ of the electromagnetic field corrected term
$(F_{\mu\nu}F^{\mu\nu})^2$. Here we further show that at this order
the term, $ F_{\mu\nu}F^{\nu\rho}F_{\rho\sigma}F^{\sigma\mu}$, has
also no contribution to the shear viscosity. In this case, the
action we are considering can  be written as
\begin{equation}
\label{eq37}
 S=\frac{1}{16\pi G}\int d^5x\sqrt{-g}\Big(
R-2\Lambda+I(F) \Big),
\end{equation}
where
\begin{equation}\label{if}
I(F)=-F_{\mu\nu}F^{\mu\nu}+\frac{(F_{\mu\nu}F^{\mu\nu})^2}{8b^2}+cF_{\mu\nu}F^{\nu\rho}F_{\rho\sigma}F^{\sigma\mu}.
\end{equation}
Here $b$ and $c$ can be arbitrary constants which represent the
effects of the higher derivative corrections of the gauge fields.
The term proportional to $1/b^2$ is the one appearing as the first
order correction of the Born-Infeld action. As the case of the AdS
Born-Infeld black holes, we consider a Ricci flat black hole
solution with electric charge in the action (\ref{eq37}). In this
case, the only nonvanishing component of electromagnetic field is
$F_{tr}$. The term $(F_{\mu\nu}F^{\mu\nu})^2$ is then just twice of
$F_{\mu\nu}F^{\nu\rho}F_{\rho\sigma}F^{\sigma\mu}$. Thus the results
concerning the corrected term only depend on the combination of the
two coefficients by $1/4b^2+c$~\cite{Anninos:2008sj}. That is, at
this order, we can obtain the black hole metric and electric field
for the action (\ref{eq37}) by replacing the coefficient $1/b^2$ in
the Born-Infeld black hole metric and electric field with the
coefficient $1/b^2+ 4c$. In the above, we have already shown that
the coefficient $1/b^2$ does not explicitly appear in the shear
viscosity. We then conclude that the term
$F_{\mu\nu}F^{\nu\rho}F_{\rho\sigma}F^{\sigma\mu}$ also will not
change the ratio of the shear viscosity to the entropy density.

\section{Conclusions and Discussions}

In this paper we calculated the ratio of shear viscosity to entropy
density in the background of AdS Born-Infeld black holes through
AdS/CFT correspondence to the first order of the Born-Infeld
parameter $1/b^2$. Motivated by \cite{Brigante:2007nu} we find that
though the higher derivative corrections to the gravity term make
the universal bound be modified, the higher derivative corrections
to the gauge fields have no change on the value of the $\eta/s$
ratio at least at the first non-trivial order.

This result may give us some hint on in what sense the lower bound
is universal. We learn that when no gravity  corrections are
considered, the ratio of $\eta/s$ is the same for various gravity
backgrounds \cite{Buchel:2003tz} and for gauge theories with nonzero
chemical potentials. When gravity corrections are added, the ratio
changes and even the lower bound could be violated, but higher
derivative corrections to the gauge fields on the gravity side do
not change the value of $\eta/s$. The difference between these two
corrections is that with the gravity correction the effective action
of gravity part is changed while in the latter case the effective
action of gravity part is still the Einstein-Hilbert form though
matter fields are coupled to gravity in various ways. This may imply
that the universality of the ratio $1/4\pi$ is just the universality
among gauge field theories which have Einstein-Hilbert gravity as
their gravity duals which can couple to arbitrary matter fields
through various proper ways. In other cases the ratio should be
mainly determined by the effective gravity part of the dual gravity
description~\cite{Brustein:2008cg}. The result of
\cite{Buchel:2008wy,{Buchel:2008ae}} might be viewed as
 evidence to support this idea, where the author found some evidence of
universality of shear viscosity at finite t'Hooft coupling. In a
recent paper~\cite{Ge:2008ni}, it has been observed that in the case
of Gauss-Bonnet gravity, the existence of chemical potentials
changes the ratio, too.

\section*{Acknowledgements} This work was supported in part by a
grant from Chinese Academy of Sciences, grants from NSFC with No.
10325525 and No. 90403029.

\end{document}